\documentclass[aps,prl,twocolumn,showpacs,preprintnumbers]{revtex4}
\usepackage{times,mathptmx,amssymb}
\usepackage{graphicx}
\usepackage{epsfig}
\usepackage{verbatim}
\usepackage{amsmath}    
\usepackage{amsfonts}
\usepackage{bbm}        

\newcommand{\eg}{{\it e.g.}}

\renewcommand{\d}{\mathrm{d}}
\newcommand{\p}{\partial}
\newcommand{\Order}{\mathrm{O}}

\newcommand{\e}{\mathrm{e}}
\newcommand{\avg}[1]{\left\langle #1 \right\rangle} 

\newcommand{\rb}[1]{{\mathbf{#1}}}  

\renewcommand{\r}{{\rb{r}}}
\newcommand{\n}{{\rb{n}}}
\newcommand{\F}{{\rb{F}}}

\newcommand{\BesselK}{{\mathrm{K}}}
\newcommand{\BesselI}{{\mathrm{I}}}

\begin{document}

\title{Violation of action--reaction and self-forces induced by
    nonequilibrium fluctuations}

\author{Pascal R. Buenzli and Rodrigo Soto} \affiliation{\mbox{Departamento
        de F\'{\i}sica, FCFM, Universidad de Chile, Santiago, Chile}\\{\rm
        (\today)}}


\begin{abstract}
	We show that the extension of Casimir-like forces to fluctuating fluids
	driven out of equilibrium can exhibit two interrelated phenomena
	forbidden at equilibrium: self-forces can be induced on single
	asymmetric objects and the action--reaction principle between two
	objects can be violated. These effects originate in asymmetric
	restrictions imposed by the objects' boundaries on the fluid's
	fluctuations. They are not ruled out by the second law of
	thermodynamics since the fluid is in a nonequilibrium
	state. Considering a simple reaction--diffusion model for the fluid, we
	explicitely calculate the self-force induced on a deformed circle. We
	also show that the action--reaction principle does not apply for the
	internal Casimir forces exerting between a circle and a plate. Their
	sum, instead of vanishing, provides the self-force on the circle-plate
	assembly.
\end{abstract}

\maketitle

In his pioneering work, Casimir \cite{casimir} showed that two metallic
plates in the electromagnetic field of vacuum attract one another due to
the restriction they impose on the quantum fluctuations of the field.
Fluctuation-induced forces exerting between macroscopic objects have since
been exhibited in a vast variety of other systems at equilibrium
\cite{kardar-golestanian}, such as critical fluids and crystal liquids in
the nematic phase, in which thermal fluctuations of long range can develop
\cite{krech,AdjariProst}. These forces are thought to play an important
role in the stability of equilibrium phases of mesoscopic particles
embedded into complex fluids. While the theoretical achievements in this
field are numerous, a direct measurement of the Casimir force in critical
fluids has only been obtained recently \cite{critcasimir}.

Matter in nonequilibrium steady states also develops fluctuations that are
generically of large amplitude and have long correlation lengths
\cite{HalperinHohenberg,OrtizdeZarate}. By analogy to the equilibrium
situation, one expects these fluctuations to induce similar forces that may
be responsible for the aggregation and/or segregation mechanisms observed
in fluids driven out of equilibrium
\cite{golestanian,voth-etal,whitesides-grzybowski}.  However, the
calculation of these forces cannot rely on the derivation of an
(equilibrium) thermodynamic potential. It is only recently that these
forces have been obtained between two planar objects immersed into
nonequilibrium driven systems \cite{bartolo-ajdari-fournier}, granular
fluids \cite{brito-soto-granular}, or reaction--diffusion systems that
violate the detailed balance \cite{brito-soto-reacdif}.

Whether at or out of equilibrium, accurate experimental measurements of
fluctuation-induced forces need to go beyond the idealized geometry of
infinitely-long plates, predominant in theory for its simplicity. Although
a long-studied topic, the proper account of the geometry dependence of
Casimir forces is a notoriously difficult problem when dealing with
nontrivial geometries \cite{balian-duplantier}. To date, the most widely
used technique by experimentalists \cite{critcasimir} relies on the
so-called Derjaguin construction \cite{derjaguin1} (proximity force
approximation) which in essence integrates the two-plate expression of the
force along the curved surfaces.

In this Letter, we show that when restricted by nonplanar objects,
nonequilibrium fluctuations can lead to additional effects not possible in
equilibrium systems. Namely, nonvanishing forces can be induced on single
asymmetric obstacles, and the action--reaction principle between two
intruders can be violated. These phenomena can have significant
consequences in experiments, as they lead to directed motion and unevenness
in the measures of the forces between two objects. Furthermore, an
unbalance of action--reaction would impeed the use of the Derjaguin
approximation where it would normally be valid at equilibrium. We found no
mention of these facts in the literature.

Since nonequilibrium systems are thermodynamically open, it is not entirely
unexpected that self-forces can appear. In fact, provided that both the
microscopic time-reversibility and space rotation-invariance symmetries are
broken, such forces have been implicitely suggested by the occurence of
sustained motions in other nonequilibrium contexts, such as in ratchets
\cite{vanDenBroeck-meurs-kawai}, Brownian motors
\cite{astumian,vanDenBroeck-meurs-kawai}, molecular motors
\cite{juelicher-ajdari-prost}, or the adiabatic piston
\cite{najafi-golestanian}. In these systems, the space asymmetry usually
lies in an external temperature gradient or in an anisotropic field
exerting on the object. More recently, however, the directed motion of an
asymmetric object immersed into vibrated granular matter has been exhibited
\cite{Constantini}. The direct calculation of self-forces that we present
here allows for a better understanding of the different effects at play in
such motions. It also makes possible the evaluation of additional stresses
exerting on asymmetrical structures in micro-devices, and could be used in
tailoring mechanisms for the self-assembly of ordered structures. The
violation of action--reaction between two intruders directly results from
the presence of self-forces. It does not seem to be systematic, however,
even in asymmetric setups \cite{antezza-etal}. Note that it prevents the
two-body forces to derive from an effective potential, in contrast to
equilibrium cases. Let us add that a violation of Newton's third law has
also been noted in depletion forces between identical spheres in a flowing
fluid \cite{dzubiella-loewen-likos}.

For illustration purposes, we exhibit here these effects with the rather
simplified nonequilibrium fluid that has been used in
\cite{brito-soto-reacdif} in a planar geometry. The applicability of the
model to nontrivial geometries can be greatly facilitated by devising an
adequate Green function formalism. We then show that the self-force
exerting on a deformed circle is indeed nonzero at second order in the
radius perturbation when dipolar deformations are considered. We also
calculate the internal forces between a circle and a plate in an asymptotic
regime where the circle's radius is small in comparison to the correlation
length while its separation to the plate is large. In this situation,
action--reaction is not satisfied as the circle-plate assembly experiences
a net self-force.

The fluctuating media is described by a reaction--diffusion fluid, whose
nonequilibrium steady state is achieved by violating the detailed balance
\cite{gardiner}. The local density $\rho(\r,t)$ of the fluid fluctuates
around a homogeneous reference density $\rho_0\equiv\avg{\rho(\r,t)}$,
where $\avg{\cdot}$ is a stochastic average. The density deviation
$\Phi(\r,t) \equiv \rho(\r,t)-\rho_0$ is assumed to satisfy, at the
mesoscopic scale, the stochastic reaction--diffusion equation
\begin{align}
	\frac{\p \Phi}{\p t} = (D \nabla^2 - \gamma) \Phi + \xi,
\label{reaction-diffusion}
\end{align}
where $D$ is a diffusion constant, $\gamma$ the reaction rate that drives
the system to local equilibrium, and $\xi(\r,t)$ is a random white noise of
correlation intensity $\Gamma$ that takes into account the fluctuations on
the reaction rates.
Eq. \eqref{reaction-diffusion} primarily describes density
fluctuations in a fluid with two reacting and diffusing chemical components
(see \cite{gardiner}), but other nonequilibrium systems
in their steady state are described by this model. The steady state fluctuations
are characterized by the bulk correlation length
$\kappa^{-1}\equiv(\gamma/D)^{-1/2}$
that can be chosen as the mesoscopic scale.

Static objects immersed in the fluid prevent any flow of matter across
their surface. Eq. \eqref{reaction-diffusion} is thus supplemented by the
non-flux condition $\n \cdot \nabla\Phi = 0$ at the objects' surface, where
$\n(\r)$ is a unit normal vector pointing outward from the fluid's
domain. In a steady state, one expects the pressure $p$ of the fluid to be
related to the density by a local equation of state $p=p(\rho(\r,t))$. This
relation is experimentally measured in a number of cases of interest, like
in driven granular media \cite{eqestadogranular}, for example. Here we only
assume that it is expandable around the reference density $\rho_0$ and that
density fluctuations stay small. The average pressure is thus modified by
the fluctuations according to $\avg{p} = p_0 + \frac{p_0''}{2}\avg{\Phi^2}$
where $p_0=p(\rho_0)$ and $p_0''=\frac{\p^2 p}{\p\rho^2}(\rho_0)$. The
total force $\F_S$ exerted by the fluid on an immersed object $S$ results
from summing this local pressure on every element $\d\sigma$ of its
surface. Since $p_0$ is a homogeneous pressure, it does not induce a force
and one has
\begin{align}
	\F_S = \frac{p_0''}{2} \int_S\!\! \d\sigma\ \n\ \avg{\Phi^2}.
\label{FS1}
\end{align}

Starting from a dynamical model like \eqref{reaction-diffusion}, Casimir
forces can be obtained by using Green functions \cite{AdjariProst}. After
an initial transient of characteristic time of order $\Order(\gamma^{-1})$,
the stationary solution of \eqref{reaction-diffusion} is
\begin{align}
	\Phi_\text{st}(\r,t) =
\int\!\!\! \d t'
	\!\!\int_\Omega \!\!\!\d\r'\, G(\r,\r',Dt\!-\!Dt')\xi(\r',t'),
\label{Phi-st}
\end{align}
where $\Omega$ is the domain occupied by the fluid.

It can be established that the temporal Fourier transform of the Green
function $G(\r,\r',\omega) = \int\!\! \d\tau\,
\e^{i\omega\tau}G(\r,\r',\tau)$ with $\tau=D t$, is, up to a factor, the
static structure factor of the fluid that enters into the force \eqref{FS1}
when evaluated at $\omega=0$:
\begin{align}
	\avg{\Phi_\text{st}(\r,t)\Phi_\text{st}(\r',t)}
	&= \frac{\Gamma}{D}\! \int\!\!\frac{\d\omega}{2\pi} \!\int_\Omega \!\!\!\d\r''
	G(\r,\r'',\omega) G(\r'',\r',-\omega) \notag\\
&=
	\frac{\Gamma}{2D}G(\r,\r',\omega\!=\!0).
\label{structfact-green}
\end{align}

The first equality in \eqref{structfact-green} follows directly from
\eqref{Phi-st} and the convolution theorem. The
second results from the differential equation satisfied by
$G(\r,\r',\omega)$ (see details in \cite{buenzli-soto-inprep}):
\begin{align}
	&(-\nabla^2+\kappa^2-i \omega)G(\r,\r',\omega) = \delta(\r-\r'),
\label{G-eq}
	\\&\n(\r) \cdot \nabla G(\r,\r',\omega)\vert_{\r\in S} = 0 \quad
	\forall \r'\in\Omega, \forall \omega. \label{G-bc}
\end{align}
In view of \eqref{structfact-green} one can omit any further reference to
$\omega$ and calculate $G(\r,\r')$ solution of \eqref{G-eq}--\eqref{G-bc}
in which $\omega$ is set to $0$ right away. This is an appreciated
simplification; $\F_S$ only depends linearly on the static Green function.

To deal with the difficulties brought about by nonplanar objects in Casimir
forces, a natural way is to use multiple scattering techniques
\cite{balian-duplantier}. If $G_0$ is the free space (unconstrained fluid)
Green function, \eqref{G-eq}--\eqref{G-bc} are also equivalent to
\cite{buenzli-soto-inprep}
\begin{align}
	G(\r,\r') = G_0(\r\!-\!\r') \!-\! \int_S\!\!\! \d\sigma_1\ G(\r,\r_1)
	\ \n_1 \!\cdot\!\nabla_1 G_0(\r_1\!-\!\r'),
\label{multiple-scattering}
\end{align}
which we abbreviate as $G=G_0 + G\star \nabla G_0$. The recursive iteration
of this integral equation expands $G$ as a series of multiple scatterings
of $G_0$ on the surface $S$. In three dimensions,
$G_0(\r)=\exp\{-\kappa|\r|\}/4\pi|r|$ and in two dimensions,
$G_0(\r)=\BesselK_0(\kappa|\r|)/2\pi$, where $\BesselK_0$ is the modified
Bessel function of order $0$.  As it is obvious from
\eqref{multiple-scattering} and the above expressions for $G_0$, the
problem of calculating the force \eqref{FS1} as it stands is ill-defined:
short-range divergences appear. They are due to the inaccuracy of the
continuous model on the microscopic scale \cite{brito-soto-reacdif}. A
``bulk'' divergence occurs when evaluating $G$ at a same point $\r$, as
well as a ``wall'' divergence once this point is approached to a
surface. The first divergence is trivial to remove: it is independent of
the immersed objects and thus consists in a homogeneous (although infinite)
pressure unable to produce a force. The wall divergence plays an important
role: it originates in the boundary condition imposed on $G(\r,\r')$ by the
immersed objects and it is integrated all along their surface in
calculating the force. The issue is then to understand how this integrated
divergence compensates itself between different sides of the objects to
yield a finite result. This compensation does not occur for any shape. To
illustrate this, we consider the Green function $G_P$ of a fluid restricted
to a half space by a plate. The condition \eqref{G-bc} on the plate can be
replaced by the addition of an ``image'' source $\delta(\r-\r'^\ast)$ on
the r.h.s. of \eqref{G-eq}, where $\r'^\ast$ is the point symmetric to
$\r'$ w.r.t. the plate. The solution hence reads
$G_P(\r,\r')=G_0(\r-\r')+G_0(\r-\r'^\ast)$. It is clear that evaluating
$(G_P-G_0)(\r,\r)$ (having subtracted the bulk divergence $G_0(\r-\r')$ as
$\r'\to\r$) at the plate's surface produces a divergent collapse of $\r$
and its image $\r^\ast$. For a smooth surface, this divergence is only
slightly modified by the curvature and one expects the total force to be
finite. By contrast, objects with sharp corners produce several images of
$\r$ (for instance, three in a right corner in 2D). They generate
additional divergences in the edges that are not likely to be compensated
(unless a symmetric corner exists on the other side of the
    object). We restrict ourselves to objects with radii of curvature
large enough to avoid such complication. The proper mathematical way of
removing the divergences is to introduce short-range (\eg, ``hard-core'')
cutoffs that are removed at the end of the calculation, similar to
classical Coulombic systems between opposite charges and at metallic
walls. Using \eqref{structfact-green}, the regularized force $\F_S$
\eqref{FS1} on the object $S$ is thus given by
\begin{align}
\F_S = \lim_{\epsilon\to 0} F_0 \kappa 
\!\int_S \!\!\d\sigma\, \n(\r)
\left[G-G_0\right](\r\!-\!\epsilon\n(\r),\r\!-\!\epsilon\n(\r)),
\label{FS-2}
\end{align}
where $F_0\equiv p_0''\Gamma/(4D\kappa)$ has the dimension of a force.  We
now apply this general framework to calculate the fluctuation-induced force
\eqref{FS-2} on two distinct systems embedded into the fluid. For
simplicity, we limit here the fluid to two dimensions, but the conclusions
also apply to three-dimensional cases.

\paragraph{Deformed circle.}
Since the force \eqref{FS-2} on a single circle must vanish by symmetry, we
deform its radius $R$ according to $R(\theta) = R + \eta s(\theta)$ (in
polar coordinates), and assume $\eta\ll \kappa^{-1}$, $R$. The general
solution of \eqref{G-eq} for a finite object is
\begin{align}
G(\r,\r') &= G_0(\r\!\!-\!\!\r')\!
+ \!\!\!\!\!\sum_{m,n\in\mathbb{Z}}\!\!\!\!\!\!
	\frac{\e^{im\theta+in\theta'}\!\!}{2\pi}
	a_{mn} \BesselK_m(\kappa \rho) \BesselK_n(\kappa \rho'), \label{G-deform}
\end{align}
where $(\rho,\theta)$ are polar coordinates for $\r$, $(\rho',\theta')$ for
$\r'$, and $\BesselK_m$ is the modified Bessel function of order $m$.  The
coefficients $a_{mn}$ satisfy $a_{mn}=a_{nm}=a_{-m,-n}^*$ to ensure that
the Green function is real and symmetric under the interchange of $\r$ and
$\r'$. They still need to be determined from the boundary condition at
$\rho=R(\theta)$. Substituting \eqref{G-deform} in \eqref{G-bc}, one can
obtain them perturbatively in $\eta$ in terms of the Fourier coefficients
of $s(\theta)$, $s_n \equiv (2\pi)^{-1}\int_{0}^{2\pi} d\theta
e^{-in\theta} s(\theta)$.

In the force \eqref{FS-2}, it must be noted that the dependence on
$R(\theta)$ is double: explicit in $\d\sigma$, ${\bf n}$, and $G$ (via
$a_{mn}$), and implicit since the Green function is evaluated on the
boundary of the deformed circle. The whole expression is expanded in $\eta$
and it is verified that the zeroth order contribution, which corresponds to
the force on the undeformed circle, vanishes. The contribution linear in
the perturbation also vanishes. Indeed, by linearity, each Fourier mode of
$s$ can be analyzed separately; the mode $n=0$ merely corresponds to a
change in the radius of the circle; the dipolar modes $n=\pm 1$ are
equivalent to a small displacement of the unperturbed circle; all remaining
modes $|n|\geq 2$ correspond to symmetric perturbations, not having any
preferred direction. The first contribution to the force thus comes from
the second order in $\eta$ and has the form
\begin{align}
\F = F_0 (\kappa \eta)^2 \sum_m f_m\ s_m\ s_{-1-m},
\label{F-circle-general}
\end{align}
where the two components of $\F$ are expressed in the r.h.s. as a complex
number. This particular combination of the (complex) Fourier coefficients
$s_n$ is necessary to ensure that the force correctly transforms as a
vector, or, equivalently in Fourier modes, as a dipole. The real
coefficients $f_m$ are nontrivial series of Bessel functions evaluated at
$R$ or $R+\epsilon$, and the limit $\epsilon\to 0$ must be taken after
summing them.

For simplicity, we choose a specific deformation that leads to a
nonvanishing force. It can be checked that $f_0=f_{-1}=0$, as expected
because the joint deformation given by $s_0$ and $s_1$ produces again a
circle that is merely expanded and translated: its self-force vanishes. The
first nontrivial cases are given by the coupling of the $n=1$ and $n=-2$
modes of $s(\theta)$. Hence considering $s(\theta) = 2s_{1}\cos(\theta) +
2s_{2}\cos(2\theta)$, we find
\begin{align}
\F = -F_0\, s_1\, s_2\ (\kappa \eta)^2 H(\kappa R)\ \hat{\bf x},
\end{align}
where $H$ is a dimensionless function whose numeric computation is
accurately compatible with $2/(\kappa R)$ in the range $0.1 \leq \kappa R
\leq 10$.

A nonvanishing self-force is therefore produced on a deformed circle at
order $\eta^2$ for the simple deformation considered here, made of a
dipolar and quadrupolar combination.

\paragraph{Circle--plate system.}
In systems at equilibrium, fluctuation-induced forces can only appear
between two or more objects. Because the global force on the total system
must vanish, such two-body forces always satisfy the action--reaction
principle. The picture is different in systems driven out of equilibrium.

Indeed, consider two objects $S$ and $S'$ immersed in the fluid. The total
force $\F_S$ on $S$ can be separated into a self contribution $\F_S^0$,
that may already be present in the absence of $S'$, and a contribution
$\F_{S\leftarrow S'}=\F_S-\F_S^0$ due to the additional asymmetry provoked
by the presence of $S'$. Denoting by $G_{SS'}$ and $G_S^0$ the Green
functions associated to the two-object and single-object $S$ systems,
respectively, one has from \eqref{FS-2} $\F_{S\leftarrow S'} \!=\!
F_0\kappa \!\int_S\d\sigma \n (G_{SS'}-G_S^0)$ and
\begin{align}
	\F_{S\leftarrow S'} + \F_{S'\leftarrow S} = \F_{SS'} - \F_S^0 -
	\F_{S'}^0,
\label{action-reaction}
\end{align}
where $\F_{SS'}$ is the global force exerting on the assembly $S\cup S'$
considered as a whole. Note that the quantity $G_{SS'}-G_S^0$ entering in
$\F_{S\leftarrow S'}$ is well-defined: the bulk divergence and the wall
divergence on $S$, present in both $G_{SS'}$ and $G_S^0$, compensate in the
subtraction. We take as definition of the action--reaction principle for
the internal (two-body) forces of such system the vanishing of the
r.h.s. of \eqref{action-reaction}. Since fluctuation-induced forces are not
additive, this vanishing will not happen in general in the presence of
self-forces, except from obvious symmetry reasons.

\begin{figure}[htb]
\epsfig{file=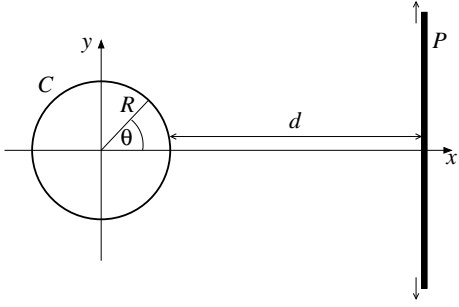,width=0.7\columnwidth}
\caption{A circle $C$ of radius $R$ and a thin, long, plate $P$ are
immersed in the fluctuating fluid. Their separation at closest point is
$d$. }
\label{fig.config}
\end{figure}

As an example, we take an assembly made of a circle $C$ of radius $R$ and a
thin and long plate $P$ in a 2D fluid. Their separation at the closest
point is $d$ (see Fig. \ref{fig.config}). The plate is taken much longer
than $\kappa^{-1}$ to avoid boundary effects. Since either object is
symmetric, $\F_C^0=\F_P^0=0$. We calculate both terms in the l.h.s. of
\eqref{action-reaction} in the regime $R\ll\kappa^{-1}\ll d$ to show that
the total force on the assembly, $\F_{CP}$, is nonzero; equivalently,
action--reaction is not satisfied in this situation.

In the multiple-scattering expansion of $G_{CP}(\r,\r)\vert_{\r\in C}$
\eqref{multiple-scattering}, the free-space correlation $G_0$ is scattered
on both surfaces $C$ and $P$. It is clear that when the separation $d$ is
much larger than the correlation length $\kappa^{-1}$, the dominant terms
in this expansion contain the least number of propagations between $C$ and
$P$. Those containing none rebuild the series of scatterings of $G_0$ on
$C$ that gives the Green function of the single circle, $G_C^0$. All other
terms contain scatterings on $P$. Since the initial and final points are
both $\r\in C$, making a scattering on $P$ necessarily involves at least
two interspace propagations, one from $C$ to $P$, and one to come back to
$C$. The next dominant terms therefore contain exactly two of the
propagators $G_0$ evaluated at points separated by at least $d$. However,
before and after these interspace propagations, any number of scatterings
from $C$ to $C$ or from $P$ to $P$ can be done. Summing them up into the
quantities $G_C^0$ and $G_P^0$, one eventually has
\begin{align}
	(G_{CP}-G_C^0)(\r,\r)\vert_{\r\in C} \stackrel{\kappa d\to\infty}{\sim}
	G_P^0 \ast \nabla G_C^0 + G_C^0 \ast \nabla G_P^0 \ast \nabla G_C^0.
	\notag
\end{align}
Explicit expressions for $G_C^0$ and $G_P^0$ are easy to obtain. The
function $G_C^0$ is straightforward to calculate from \eqref{G-deform} with
the result $a_{mn}=-(\BesselI_m'(\kappa R)/\BesselK_m'(\kappa
R))\delta_{n,-m}$, where $\BesselI_m'$ and $\BesselK_m'$ are the
derivatives of the modified Bessel functions of order $m$.

We shall here only state the result of an asymptotic analysis of
$\F_{C\leftarrow P}$ and $\F_{P\leftarrow C}$, based on small-$\kappa R$
expansions and steepest-descent values of integrals as $\kappa d\to\infty$
(explicit calculations, including in other regimes, will be developed in
\cite{buenzli-soto-inprep}). On the $x$ axis, the forces exerting on the
circle and the plate in the regime $R\ll\kappa^{-1}\ll d$ are found to be
\begin{align}
	F_{C\leftarrow P}\ \sim\ - F_0 \frac{\sqrt{\pi} (\kappa R)^2 \e^{-2\kappa
	d}}{\sqrt{\kappa d}}, 
\quad F_{P\leftarrow C} \sim -\frac{3}{2}F_{C\leftarrow P}.
\end{align}
Clearly, action--reaction is not satisfied. Furthermore, the circle--plate
assembly experiences a nonzero global self-force along $x$ of the same
order: $F_{CP} \sim \frac{1}{2} F_0 \frac{\sqrt{\pi} (\kappa R)^2
    \e^{-2\kappa d}}{\sqrt{\kappa d}}$.

In conclusion, we have shown that in nontrivial geometries the extension of
Casimir-like forces to fluctuating fluids driven out of equilibrium must
take into account two interrelated phenomena forbidden at equilibrium: the
possibility that a self-force may be induced on a single asymmetric rigid
body, and that the action--reaction principle for the forces between two
objects may be strongly violated. The latter fact impeeds that an effective
interaction potential holds in nonequilibrium. Its occurence would prevent
the use of the Derjaguin approximation. As the magnitude of this violation
can be of the same order as the internal forces, special care should be
taken when obtaining these forces in experiments or simulations.

The complexity of dealing with nonplanar objects has been overcome here by
considering a very simple model for the nonequilibrium fluid and by
devising a Green function and multiple scattering technique. Clearly, to
allow quantitative comparison with real fluids (such as colloidal
solutions, dusty plasmas, etc.), one would need to refine the initial
model.

The presence of self-forces leads to directed motion if the objects are let
free to move as in the case of ratchets. Self-forces can lead to
arrangements of composites of intruders that could be tailored to produce
microdevices by self-assembling.  Their dynamical properties, however, need
a more thorough analysis, for their motion will affect the fluid's
fluctuations and a self dynamical interaction could appear.

We would like to acknowledge R. Brito for fruitful discussions.
This work is supported by the grants {\em Fondecyt} $1061112$, $3070037$,
{\em Fondap} 11980002 and {\em Snsf} PBEL$2$-$116909$.

\end{document}